\documentstyle[namedreferences,epsfig]{kluwer}

\newcommand{\ra}{\rightarrow}
\newcommand{\et}{{\em et al.}}
\newcommand{\ie}{{i.e.}}

\begin{opening}
\title{Frequency control in synchronized networks of inhibitory neurons}

\author{Carson C. \surname{Chow}}
\institute{Department of Mathematics and Center for BioDynamics,
Boston University, Boston, MA 02215\footnote{After September 1998: 
Dept. of Mathematics, University of Pittsburgh,
Pittsburgh PA  15260}} 
\author{John A. \surname{White}}
\institute{Department of Biomedical Engineering and Center for BioDynamics,
Boston University, Boston, MA 02215}
\author{Jason \surname{Ritt}}
\author{Nancy \surname{Kopell}}
\institute{Department of Mathematics and Center for BioDynamics,
Boston University, Boston, MA 02215}

\date{\today}
\end{opening}

\begin{document}

\maketitle

\begin{abstract}

We analyze the control of frequency for a synchronized inhibitory
neuronal network.  The analysis is done for a reduced membrane model
with a biophysically-based synaptic influence.  We argue that such a
reduced model can quantitatively capture the frequency behavior of a
larger class of neuronal models.  We show that in different parameter
regimes, the network frequency depends in different ways on the
intrinsic and synaptic time constants.  Only in one portion of the
parameter space, called `phasic', is the network period proportional
to the synaptic decay time.  These results are discussed in connection
with previous work of the authors, which showed that for mildly
heterogeneous networks, the synchrony breaks down, but coherence is
preserved much more for systems in the phasic regime than in the other
regimes.  These results imply that for mildly heterogeneous networks,
the existence of a coherent rhythm implies a linear dependence of the
network period on synaptic decay time, and a much weaker dependence on
the drive to the cells.  We give experimental evidence for this
conclusion.

\end{abstract}

\keywords{gamma oscillations, hippocampus, interneurons, synchronization}

\section{Introduction}

Coherent neuronal oscillations have been implicated widely as a
correlate of brain function~\cite{gray94,llinas93}.  However, our
understanding of the mechanisms underlying such activity is still in
its infancy. (See \citeauthor{jeff96}~(\citeyear{jeff96}) and
\citeauthor{ritz97}~(\citeyear{ritz97}) for recent reviews).  It has
been demonstrated 
experimentally~\cite{whit95,jeff96},
computationally~\cite{wang92,whit95}, and
analytically~\cite{vrees94,gerstner96,terman} that mutual inhibition
can generate stable synchronous oscillations.  Previous analytical
work mostly concentrated on the mechanisms responsible for
synchronization.  Here we address the mechanisms for controlling the
network frequency and the interplay between frequency and synchronization.

Previously~\cite{white97}, we showed that synchronization of
inhibitory networks can occur in a wide range of frequencies for
homogeneous neurons, but in the presence of heterogeneity
network synchronization is very fragile in some parameter regimes.
We have identified a pair of parameter regimes,
denoted `phasic' and `tonic'; in the former some coherence is
maintained in the presence of mild heterogeneity, while in the latter
it is lost.  In this paper, we relate these regimes to the parameters that
determine the frequency of the network when it is synchronized.

There are three important time
constants in the network dynamics.  One is an intrinsic time constant of the
uncoupled cell, dependent on the conductances and the capacitance
of the membrane.  The second is the decay time of the inhibition.  The
third, the network period, depends on the other two time
scales and other parameters, notably the synaptic conductance and 
injected current.
The aim of this paper is to understand how the first and second time
constants influence the third.  

We show that there are three asymptotic regimes, in which the network
period varies differently as the inhibitory time constant is
changed. Two of these are the tonic and phasic~\cite{white97} regimes
mentioned above and the third we call `fast'.  In the tonic regime,
the period is small compared with both the intrinsic time scale and
the synaptic decay time.  In this regime, we show that the network
frequency is only weakly dependent on the inhibitory decay time;
indeed, the frequency is affected mainly by the average amount of
inhibition, not by the time course of that inhibition.  In the phasic
regime, the intrinsic membrane time scale is small compared with
the network period and the synaptic decay time.  In this regime, we
show that the network period is proportional to the decay time, with
the proportionality constant a function of other network parameters.
In the fast regime, the synaptic decay time is short compared to
the network period.  In this regime, the period is dominated by the
intrinsic time constant.  We give constraints on the network
parameters (intrinsic membrane time scale, synaptic decay time,
synaptic conductance and applied current) for the network to fall into
each of the three asymptotic regimes.

The three regimes can be related to the behavior of the network in the
presence of mild heterogeneity , as shown numerically in White
\et~(\citeyear{white97}).  In the fast regime, the synaptic influence
acts as a brief inhibitory pulse which has been shown to lead to
stable anti-synchronous oscillations
\cite{friesen94,perkel74,skinner94,vrees94,wang92}.  In the phasic and
tonic regimes, synchrony through inhibition is fairly robust for
homogeneous networks of neurons over a wide range of network
frequencies~\cite{vrees94,gerstner96}.  However,
it is very fragile when even mild heterogeneity is included
\cite{golomb93,wang96,white97}.  We showed that the loss of coherence
of the network happens by different mechanisms in the tonic and phasic
regimes \cite{white97}.  In the tonic regime, mildly heterogeneous
networks are effectively de-coupled and exhibit asynchrony (loss of
phase relationships among the cells); in the phasic regime, the
network can lose some coherence via suppression, in which the slower cells
receive enough inhibition to prevent their firing.

Our analytical work uses a reduced membrane description for the neuron
with a biophysically-based synaptic model.  The analysis is for a
homogeneous network; when considering the frequency of a synchronous
solution, we can regard the network as a single self-inhibited neuron.
We show that in the phasic regime, the synaptic current dominates and
the actual intrinsic membrane dynamics are not as important for the
firing frequency.  In the tonic regime, the intrinsic dynamics do
become more important.  We give a comparison between our
analytical estimates for the period and those obtained numerically
for conductance-based neuron models.

The work in this paper, in conjunction
with~\citeauthor{white97}~(\citeyear{white97}) shows that, in
the presence of mild heterogeneity, the control of frequency is
strongly tied to the time scale of the mechanism that produces the
synchronization.  That is, in order to have coherence for a mildly
heterogeneous system, the latter must be in the phasic regime, which
then implies that the frequency is controlled by the time constant of
the inhibitory decay.  It was found experimentally and numerically
that the frequency of coherently firing interneurons in the CA1 region
of the hippocampus is strongly dependent on the decay time of the
inhibitory synapse~\cite{whit95,jeff96,traub96,wang96}.  The analytical
work given in this paper clarifies the reasons for this frequency
behavior as well as the remark found in many papers that the
inhibitory decay time can be a critical factor in the determination of
the network frequency~\cite{destexhe93,skinner93,kopell}.

\section{Neuron model}

We consider the frequency control of so called Type I
neuronal dynamics~\cite{hansel95,bard96} and examine 
simplifications that would make it more amenable to
analysis.  These neurons are distinguished by the fact
that they have positive {\em phase response
curves} (PRC)~\cite{hansel95}. \ie~a positive depolarizing current always
advances the time of the next spike.
It has been shown recently that neurons which
admit very low frequency oscillations near the critical applied
current are Type I~\cite{bard96} although the converse is not
necessarily true. We note that the type of bifurcation to firing of
the neuron is not important for our 
analysis. Type I neurons
have been used to represent inhibitory interneurons
in the hippocampus~\cite{traub96,wang96,white97}.

\begin{figure} 
\centerline{\epsfig{figure=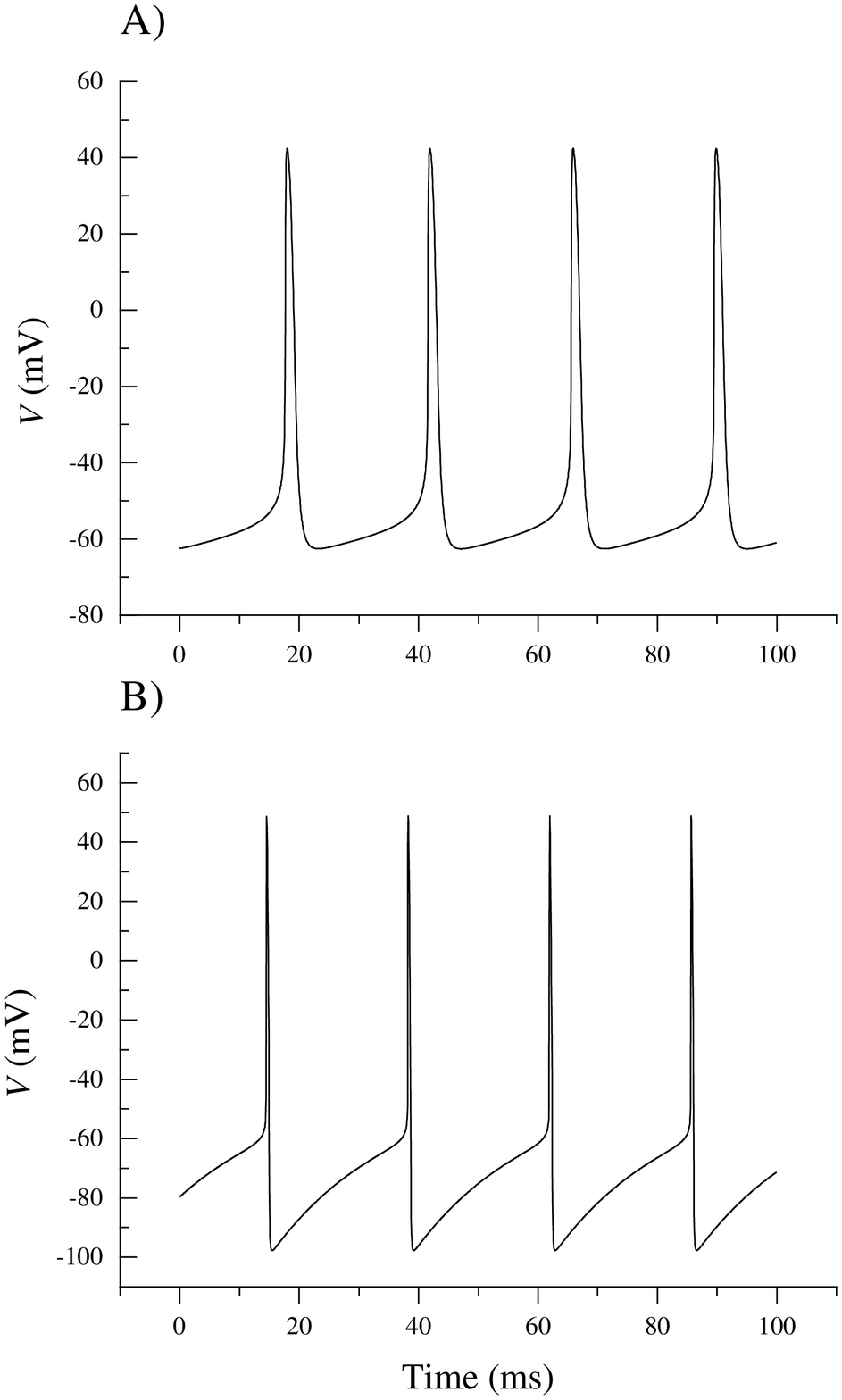, height=7.in,bbllx=14pt,bblly=9pt,bburx=597pt,bbury=784pt, clip=}}
\caption{Example voltage traces of spikes for the A)
White~\et~(1998) interneuron and B) reduced
Traub and Miles model of Ermentrout and Kopell~(1998).}
\label{fig:spike}
\end{figure}

We consider an inhibitory network of single-compartment neurons
with membrane 
potential dynamics of the form~\cite{rinzel,bard96,hansel95,wang96,white97} 
\begin{equation}
C\frac{dV_i}{d\tilde{t}}= \tilde{I}_i- \sum I_{\rm ion} - I_s,
\label{mem1}
\end{equation}
where $\tilde{I}_{i}$ is an applied current, $I_{\rm ion}$ are the
ionic currents responsible for spike generation and recovery and $I_s$
is the synaptic current induced by the spikes of other neurons coupled
pre-synaptically.  The ionic currents $I_{\rm ion}$ are functions of
potential and the dependent dynamical variables, which are in turn
governed by a system of differential equations.  In the Appendix we
give the equations for two conductance-based neuron
models~\cite{white97,ermenkopell98}, which are Type I or at least very
close to Type I (data not
shown).  Figure~\ref{fig:spike} shows example voltage traces for the
two models.  The coupling between the neurons is exclusively through
chemical synapses represented by the synaptic current which takes the
form $I_s=\tilde{g} S(\tilde{t})(V-V_s)$, where
\begin{equation}
\frac{dS}{d\tilde{t}} = \tilde{\alpha} F(V_{pre})(1-S)-\tilde{\beta} S,
\label{synapse}
\end{equation}
$\tilde{\alpha}$ and $\tilde{\beta}$ are respectively the
synaptic rise and decay rates, $V_{pre}$ is the membrane potential of
the pre-synaptic neuron,
and  $F(V)=1/(1+\exp[-V])$.  We will
often consider the synaptic decay time
$\tilde{\tau}\equiv\tilde{\beta}^{-1}$.

We approximate the full conductance-based dynamics with
integrate-and-fire dynamics where the firing of the neuron is
represented by the resetting of the membrane potential whenever it
crosses a threshold.   Our justification for using this
approximation hinges on three considerations: 1) both
the conductance-based and integrate-and-fire models
are Type I (in the sense of positive PRC), 2) the action potentials
(spike widths) are narrow compared to their typical spiking period so
the frequency is dominated by the membrane recovery time, and 3) the time
scale for spike generation is very fast compared to the recovery time
so an effective threshold for spiking can be defined.  The first point
was verified by observing that the measured PRC of the
conductance-based models near
the bifurcation to firing is positive. 

We model the approach to threshold with a simple
passive decay to obtain dynamics governed by a single equation of the
form:
\begin{equation}
C\frac{dV}{d\tilde{t}}\simeq \tilde{I} - g_m(V-V_r) - \tilde{g}
S(\tilde{t})(V-V_s), 
\label{if1}
\end{equation}
where $g_m$ is an effective membrane recovery conductance, $V_r$ is an
effective membrane reversal potential and $V_s$ is a synaptic reversal
potential.  $V(t)$ is reset to $V_0$
whenever it reaches the threshold potential $V_T$.  
The passive decay to threshold is a very good approximation for some
neuronal models such as the reduced Traub and Miles
model given in the Appendix~\cite{ermenkopell98}.  On the other hand,
we will show that even 
when the passive decay is not a good approximation to the slow
dynamics of the neuron model, it can still adequately describe the
frequency behavior, especially in the phasic regime where the synaptic
current dominates.

The synaptic current $S(\tilde{t})$ is generated from the spikes of
pre-synaptic neurons.  This 
must be emulated in the reduced model (\ref{if1}).  Here we consider
$S(\tilde{t})$ to be an arbitrary time dependent function.  In
Sec.~\ref{sec:syn}, we analyze some biophysical synaptic models in detail and 
explicitly derive the time course of $S(\tilde{t})$ in response to a
pre-synaptic spike.

To simplify the analysis, we rescale Eq.~(\ref{if1})
so that only dimensionless parameters remain. 
The voltage can be rescaled via 
$v = (V-V_0)/(V_T-V_0)$, so that
the reset potential is at $v=0$ and the threshold is at $v=1$. We
define an intrinsic membrane decay time 
\begin{equation}
\tau_m\equiv C/g_m
\end{equation}
and rescale time by $t=\tilde{t}/\tau_m$.
This scaling takes the membrane time scale to 1,
leading to the dimensionless equation
\begin{equation}
\frac{dv}{dt}= I - [1 + \gamma S(t)]v -  g S(t), 
\label{re}
\end{equation}
where
\begin{equation}
I=\frac{\tilde{I}+ I_r}{I_T}, \qquad g= \frac{\tilde{g}}{g_T}, 
\qquad \gamma = \frac{\tilde{g}}{g_m},
\label{transform}
\end{equation}
with
$I_T=g_m (V_T-V_0)$, $I_r= g_m(V_r-V_0)$, and $g_T = I_T/(V_0-V_s)$.
The neuron is said to fire
each time $v(t)$ reaches 1, whereupon it immediately resets to  
$v=0$.  It is important to note that this does not restrict
$v(t)$ to nonnegative values away from $t=0$.

An even simpler model is
\begin{equation}
 \frac{dv}{dt}=I -  v - g S(t).
\label{re2}
\end{equation}
which follows from Eq. (\ref{re}) if $\gamma << 1$.
We refer to the reduced model given by Eq.~(\ref{re}) as model 1 and
the one given by Eq.~(\ref{re2}) as model 2.
The two  models differ in
that the synapse acts solely as a forcing function in model 2, but
also affects the membrane decay rate in model 1. 
As we will show, the frequency behavior of the two models is
qualitatively
similar
even if $\gamma$ is not small.
We will focus most attention on model 2 in our analysis.  The 
synchronization tendencies of model 2 for slow synapses has been studied 
thoroughly \cite{vrees94,hansel95,gerstner96,chow97}.

\section{Synaptic Model}\label{sec:syn}

In conductance-based neuron models, the post-synaptic current is
initiated by 
a pre-synaptic spike.  In the reduced model, the spikes have been
eliminated so this process must be modeled.  We do so by
deriving the synaptic time course for Eq.~(\ref{synapse})
explicitly for a stereotypical pre-synaptic spike.  
As we will show below, the spike response can be
described by a time dependent recursive forcing function of the form
\begin{equation}
S(t) \ra  a(t-t_l) S(t) + S_f(t-t_l),
\label{recur}
\end{equation}
where $S(t)$ is the synaptic gating variable to be used in
Eq.~(\ref{re}), $t_l$ gives the spiking times of the
pre-synaptic neuron, $S_f(t)$ is the stereotypical post-synaptic
response, and $a(t)$ is a `memory' function. (Note that all time and
rate parameters
have been rescaled by $\tau_m$ in this section).
We divide synapses
into two types -- {\em saturating} and {\em nonsaturating}. If the
synaptic variable $S(t)$ increases without bound as the rate of firing 
of the pre-synaptic cell approaches infinity then we call this a
nonsaturating synapse.  If however, $S(t)$ saturates to a fixed value
as the firing rate approaches infinity then we call this a saturating
synapse.  The behavior of the function $a(t)$ determines the type of
the synapse. 

We first derive the recursion relation explicitly for
a simple nonsaturating synaptic model given by
\begin{equation}
\frac{dS}{dt}= -S/\tau + \sum_{l=0}^{N} \delta(t-t_l),
\label{synns}
\end{equation}
where $t_l$ denotes the times of the pre-synaptic spikes.
Synapses of this form have been considered in
previous models~\cite{abbott93,tsodyks93,vrees94,hansel95}.  Integrating 
Eq.~(\ref{synns}) results in 
\begin{equation}
S(t) = S(0)e^{-t/\tau} + \sum_{l=0}^N e^{-(t-t_l)/\tau},
\end{equation}
where we begin the integration at $t=0$.
This sum is equivalent to the recursion relation 
\begin{equation}
S(t) \ra S(t) + e^{-(t-t_l)/\tau}.
\label{vvrec}
\end{equation}
Here $a(t)=1$ and $S(t)$ increases without bound as the frequency
increases.

Now consider the synaptic
model in the Appendix. We will show that this is a saturating
synapse. When the   
pre-synaptic neuron fires, $F(V_{\rm pre})$ in Eq.~(\ref{synapse})
rises from near zero to a value near unity, then returns to zero.  The
precise shape is determined by the temporal characteristics of the
action potential and $F(V)$. Our analysis is similar to that
of~\citeauthor{destexhe94}~(\citeyear{destexhe94}).
We assume that the shape of $F(V)$ can be
approximated by a square pulse with a width of $\Delta t$, given by
the time the membrane potential remains above zero during an action
potential (\ie~width of the spike).  We consider a train of square
pulses arriving at times 
$t_l$, with $t_l + \Delta t < t_{l+1}$ \ $\forall\,l$ (\ie~no
overlapping spikes).  The synaptic response is then found from
\begin{equation}
\frac{dS}{dt}=\alpha Q(t)(1-S)-\beta S,
\label{syn2}
\end{equation}
\begin{equation}
Q(t) = \sum_{l=-\infty}^{\infty} H(t-t_l),
\end{equation}
where $H(t)$ is a square pulse of unit height with rise at $t=0$ and fall 
at $t= \Delta t$.

Given $l$, we define
\begin{equation}
R(t) \equiv \alpha \int_{t_l}^{t} Q(t') dt' 
       = \left\{\begin{array}{ll} 
                \alpha (t-t_l), & t_l \le t \le t_l+\Delta t\\
                \alpha \Delta t, &  t_l + \Delta t < t < t_{l+1}
                \end{array}
       \right.  \end{equation} and then integrate~(\ref{syn2}) from
$t_l$ to $t<t_{l+1}$ to yield \begin{eqnarray} S(t) &=&
e^{-R(t)-\beta t} \left[S(t_l) e^{\beta t_l} +
\int_{t_l}^t \alpha Q(t') e^{R(t')+\beta t'}
dt'\right] \\ &=& \left\{ \begin{array}{ll}
e^{-\alpha(t-t_l)}(S(t_l)e^{- \beta(t-t_l)}) +
\left(\frac{\alpha}{\alpha+\beta}
\right)\left(1-e^{-(\alpha + \beta)(t-t_l)}\right), &
t_l \le t \le t_l+\Delta t \\ e^{-\alpha \Delta t} (S(t_l) e^{
-\beta(t-t_l)}) + \left(
\frac{\alpha}{\alpha+\beta}\right) \left(1 -
e^{- (\alpha+\beta) \Delta t}\right)e^{-
\beta(t-t_l-\Delta t)}, & t_l + \Delta t < t < t_{l+1}
\end{array}\right.  \label{S} \end{eqnarray}
Consider $S(t)$ at the time of the next spike, $t=t_{l+1}$.  If the
spike does not occur, $S(t)$ will continue to decay exponentially with
rate $\beta$, \ie
\begin{equation}
\begin{array}{ll}
S(t)=S(t_{l+1})\exp(-\beta(t-t_{l+1})), &  t_{l+1} < t.
\end{array}
\label{nospike}
\end{equation} 
We can thus rewrite
Eq. (\ref{S}) as a recursion relationship with the spike at $t_{l+1}$ as
\begin{equation}
S(t)\ra a(t-t_{l+1}) S(t)+S_f(t-t_{l+1})
\label{rec}
\end{equation}
where
\begin{equation}
a(t) = \left\{\begin{array}{ll}
e^{-\alpha t}, 
& 0\le t\le \Delta t,\\ 
e^{-\alpha \Delta t}, 
& \Delta t < t,
\end{array} \right.
\label{f}
\end{equation}
and
\begin{equation}
S_f(t) = \left\{\begin{array}{ll}
(\frac{\alpha}{\alpha+\beta})
(1-e^{-(\alpha+\beta)t}), 
& 0\le t\le \Delta t, \\ 
(\frac{\alpha}{\alpha+\beta})
\left(1 - 
e^{-(\alpha+\beta) \Delta t}\right)e^{-\beta
(t-\Delta t)},  
& \Delta t < t.
\end{array} \right.
\label{Sf}
\end{equation}
The $S(t)$ appearing in the right hand side of Eq. (\ref{rec}) is
understood to be the function that would occur in the absence of a new
spike, Eq. (\ref{nospike}).

In contrast to Eq. (\ref{vvrec}), there is a multiplicative
factor $a(t)$ which `damps' away 
the past.  For $a(t)<1$, $S(t)$ always saturates to a finite value.
$S_f(t)$ and $a(t)$ have discontinuous first derivatives because a
square pulse was used as input, but in general the functions will be
smooth.

Depending on the
values of the three parameters $\alpha$, $\beta$ and
$\Delta t$, the synaptic time course can take on many shapes.  Here we
are concerned with narrow spikes (small $\Delta t$) and slowly
decaying inhibition ($\beta << \alpha$). 
If the rise time is very fast, we can explicitly take the double limit
$\alpha\ra\infty$, $\Delta t \ra0$, $\alpha\Delta t=c$,
where c is a constant, to get 
\begin{equation}
S_f(t)= (1-e^{-c})e^{-\beta t}.
\label{histupdate}
\end{equation}
and $a=e^{-c}$.

The constant $c$ is the ratio of the spike width to the rise time of
the synaptic current, and determines the contribution from the past to
the current synaptic response.  If $c$ is large enough we can
reasonably ignore the past and use the approximation $a\simeq 0$ and
$S_f(t)\simeq\exp(-\beta t)$.  Then the synaptic update
function to be used in Eqs.~(\ref{re}) and (\ref{re2}) takes the form
\begin{equation}
S(\tilde{t})\rightarrow  e^{-(t-t_l)/\tau},
\label{synupdate}
\end{equation}
where
$\tau=\tilde{\tau}/\tau_m$ is the rescaled synaptic decay time.
Equation (\ref{synupdate}) gives the recursion relation for the synapse
where $a(t)=0$ and
is valid in the limit where the rise time
of the synaptic gating variable is very fast compared to the spike
width.  We note that a nonzero rise time is very important for
synchronization~\cite{vrees94,hansel95,gerstner96,terman,chow97} but
it is not as important for frequency determination.

\section{Period of a synchronized network}\label{freq}

In this section we will derive the form of the network period
analytically for model 2 [Eq.~(\ref{re2})] in
terms of a transcendental relation for saturating and non-saturating
synapses. 
We will also analyze the differences that would
result if we considered model 1 [Eq.~(\ref{re})].  In general the same
qualitative behavior holds 
between the two reduced models.
To calculate the period for the reduced models, we assume
the neuron spikes repeatedly with a period $T$ (the
frequency is defined as $f=T^{-1}$).  We then
integrate the membrane equation from the time of the last spike to the
time of the next spike.  The period is obtained self consistently by
imposing the constraint $v(T)=1$.

Suppose the neuron last fired at $t=0$, then
for model 1 the membrane voltage obeys
\begin{equation}
v(T)=1=e^{-\mu(T)}\int_0^T e^{\mu(t')}[I- g S(t')] dt'
\label{v2}
\end{equation}
where (for model 1)
\begin{equation}
\mu(t)=\int_0^t[1 + \gamma S(t')] dt'.
\label{mug}
\end{equation}
For saturating synapses, $S(t)= \exp(-t/\tau)$, which implies that
\begin{equation}
\mu(t)=  t + \gamma (1-e^{-t/\tau})\tau.
\label{mu}
\end{equation}
For model 2, this simplifies to $\mu(t) = t$.

Equation (\ref{v2}) gives the membrane dynamics of the current spiking
cycle in response to synaptic inputs from all the previous cycles.
We can evaluate the
combined synaptic input from all of the previous spikes if we assume
periodic firing.  For the
neuron spiking in the past 
at times $t=-lT$, $l=0,1,2,\dots$, we can re-express the recursion
relation (\ref{recur}) as the sum
\begin{equation}
S(t)=\sum_{l=0}^\infty a^l S_f(t+lT).
\end{equation}
For the saturating synaptic model, $a=\exp(-\alpha \Delta t)$ and
$S_f$ is given in Eq.~(\ref{histupdate}).
This is a geometric series and can be summed to give
\begin{equation}
S(t)=(1-a) \frac{e^{-t/\tau}}{1- a e^{-T/\tau}}.
\label{syn3}
\end{equation}

For model 2 (using $\mu(t) = t$) we can integrate
(\ref{v2}) explicitly using (\ref{syn3}) to obtain
\begin{equation}
v(T) = 1 = I(1-e^{- T})-\frac{g \tau (1-a)(e^{-T/\tau} -
e^{- T})}{(\tau - 1)(1-a e^{-T/\tau})}. 
\label{transfull}
\end{equation}
Equation~(\ref{transfull}) is a transcendental relation that determines the
spiking period and hence frequency of the synchronous network. 
If we can completely ignore the past (\ie~use synaptic model
(\ref{synupdate})) then $a \sim 0$ and we obtain
\begin{equation}
v(T) = 1 = I(1-e^{- T})-\frac{g \tau }{\tau - 1}(e^{-T/\tau} -
e^{- T}).
\label{trans}
\end{equation}
For simplicity, we will conduct our analysis on this equation.

For nonsaturating synapses, we can again calculate $S(t)$ for
periodic firing as we did for the previous case.  Here, we find
\begin{equation}
S(t)=\frac{e^{-t/\tau}}{1-e^{-T/\tau}}.
\label{syn4}
\end{equation}
For model 2 (using $\mu(t) = t$) we can integrate
(\ref{v2}) explicitly using (\ref{syn4})
to obtain
\begin{equation}
v(T) = 1 = I(1-e^{- T})-\frac{g \tau (e^{-T/\tau} -
e^{- T})}{(\tau - 1)(1-e^{-T/\tau})}. 
\label{transns}
\end{equation}

\section{Frequency Regimes}

In this section we describe in detail the three asymptotic
regimes discussed in the Introduction.
We derive the parameter ranges in which each regime holds,
and the dependence of the network period on the inhibitory
decay time within each regime.

\subsection{Tonic Regime}
The Tonic regime occurs where
\begin{equation}
T<<1, \qquad T <<\tau. 
\label{Tcond}
\end{equation}
Note that unity in (\ref{Tcond}) corresponds to the intrinsic
time scale of the membrane, which has been scaled away in
our normalization, and $\tau$ represents the ratio
of the synaptic decay time to the membrane time scale.
The relationship in (\ref{Tcond}) allows us to expand the 
exponentials in Eq.~(\ref{trans}) to linear
order in $T$ and $T/\tau$ to obtain for saturating synapses
\begin{equation}
T \simeq (I-g)^{-1}.
\label{TI}
\end{equation}
To derive the condition which must be satisfied for this regime 
we impose our original assumptions (\ref{Tcond}) on
Eq.~(\ref{TI}).  This leads to the 
condition
\begin{equation}
(I-g)^{-1} << \min [\tau, 1]
\label{Tdef}
\end{equation}
Condition
(\ref{Tdef}) can be considered to be the definition of the tonic
regime.
A strong applied current (large $I$) and a
weak synapse (small $g$) suffice to satisfy it.

In the tonic regime, the synaptic decay time does not influence the
period, as can be seen from Eq. (\ref{TI}). 
The synapse only affects the
period through the average 
amount of inhibition the neuron receives over the course of one
period.  We see this immediately by replacing $S(t)$ in
Eq.~(\ref{v2}) by its average $\langle S(t) \rangle$.
After integrating, we obtain the relation
\begin{equation}
1  \simeq [I-g \langle S(t) \rangle] (1-e^{-T}),
\label{avgTrans}
\end{equation}
where the time average of the synaptic input $S(t)$ is defined as
\begin{equation}
\langle S(t)\rangle = \frac{1}{T} \int_0^T S(t') dt'.
\label{avgS}
\end{equation}
In the tonic regime to leading order this leads to the simple result of
$\langle S(t) \rangle \sim 1$ for saturating synapses
and $\langle S(t) \rangle \sim \tau/T$ for nonsaturating synapses
(using Eq.~(\ref{transns})).
If we substitute $\langle S(t) \rangle$ into  Eq.~(\ref{avgTrans}) 
and linearize in $T$ and $T/\tau$ (using (\ref{Tcond})), we obtain
the same period as given by Eq.~(\ref{TI}) for saturating synapses.
For nonsaturating synapses, we find the period obeys
\begin{equation}
T \simeq I^{-1}(1+g\tau).
\label{TIns}
\end{equation}
Unlike the saturating case there is linear dependence on
$\tau$.  However, since $g/I$ is small in this regime this dependence
is small.

The inclusion of the synaptic contribution to the decay rate in
model~1 does not alter the results.   
In the tonic regime, $T << \tau$ so we can expand Eq.~(\ref{mu}) to
obtain  
\begin{equation}
\mu(t) \simeq  t + \gamma t\equiv r t.
\label{muexpand}
\end{equation}
The inclusion of the synaptic contribution to the rate
only changes the effective passive decay rate from 1 to $r$.  Using
this in Eq.~(\ref{trans}) and expanding to linear order yields the the
same period as given in Eq.~(\ref{TI}).

\subsection{Phasic Regime}
The phasic regime occurs where 
\begin{equation}
1<< T,\qquad 1<< \tau.
\label{condP}
\end{equation}
This implies that $T>>T/\tau$ which
allows us to ignore  $e^{-T}$ with
respect to $e^{-T/\tau}$ and 1.  In this regime the neuron fires
at a frequency which is slow but on
the order of  $\tau^{-1}$.  Applying condition (\ref{condP}) to
Eq.~(\ref{trans}) yields
\begin{equation}
1=I - \frac{g \tau}{\tau-1} e^{-T/\tau}.
\end{equation}
for saturating synapses. Solving for $T$ gives
\begin{equation}
T = \tau \ln\left[\frac{g\tau}{(\tau-1)(I-1)}\right].  
\label{TD}
\end{equation}
Equation~(\ref{TD}) shows that in this regime the period is
approximately proportional to $\tau$ and logarithmically dependent on
the other parameters.  We called this regime the phasic regime in
White \et~(1997) because the synaptic variable $S(t)$ is dominated by
a phasic component with a period proportional to $\tau$.  To derive
the condition on the parameters for the network to be in the phasic
regime, we apply the condition $T>>1$ to Eq.~(\ref{TD}).  We then have
the condition
\begin{equation}
\ln\left[\frac{g\tau}{(\tau-1)(I-1)}\right] >> \tau^{-1}.
\label{Pdef}
\end{equation}
Thus, the phasic regime is obtained for $g$ large compared to $I-1$.
We should note that $I>1$ is the condition required for firing and
thus must always be satisfied.

We  can  compare 
condition (\ref{Pdef}) to condition (\ref{Tdef}) for the tonic
regime. 
Both the phasic and the tonic regimes require fairly large
$\tau$ but it is the relationship between $g$ and $I$ that
distinguishes the two regimes.  The phasic regime is attained by a strong
synapse 
and a weak applied current and the tonic regime is attained by the
opposite. 

For nonsaturating synapses using Eq.~(\ref{transns}) the period takes
the form 
\begin{equation}
T = \tau \ln\left[\frac{g\tau + (\tau-1)(I-1)}
{(\tau-1)(I-1)}\right].  
\label{TDns}
\end{equation}
The period is again proportional to $\tau$ but the logarithmic factor
is changed.

The period is modified for model~1 but remains
proportional to $\tau$.  From Eq.~(\ref{mu}) we can see that 
$\mu(t)$ is bounded by $r t > \mu(t) > t$, where $r=1 + \gamma > 1$.
Hence the period is bounded between (\ref{TD}) and
\begin{equation}
T=\tau\ln\left[\frac{r g \tau}{(r\tau-1)(I-r)}\right].
\end{equation}
Thus,  $T\propto \tau$ but the proportionality
constant is changed by a logarithmic factor.  Here $I>r$ must be
satisfied
for the neuron to fire.

\subsection{Fast Regime}
The phasic and tonic regimes are applicable for slow synapses, where
the synaptic decay time is comparable to or slower than the period.
Here, we investigate the fast regime where the synapse is very fast
compared to the period; \ie
\begin{equation}
\tau<< T.
\label{condF}
\end{equation}
The fast regime can occur for a range of $\tau$. 

We first consider $\tau<<1$ so that
we can ignore  $e^{-T/\tau}$ with
respect to $e^{-T}$.  
In this case Eq.~(\ref{trans}) for saturating synapses becomes
\begin{equation}
1\simeq I(1-e^{- T}) - g\tau e^{- T},
\end{equation}
where we have also expanded to linear order in $\tau$.  From this
we obtain
\begin{equation}
T= \ln\left[\frac{g\tau+I}{I-1}\right].
\label{T3}
\end{equation}
Note that the network period is a logarithmic
function of the parameters.
Thus, in this regime, the period is dominated by the membrane 
time constant, which is implicit in our scaling.
For this regime to be valid we must impose condition (\ref{condF}) on
(\ref{T3}).  Thus we have the condition
\begin{equation}
\ln\left[\frac{g\tau+I}{I-1}\right]>>\tau.
\label{Fdef}
\end{equation}
This can be satisfied for $I\sim 1$ for arbitrary $g$ or for $g$ large
enough and arbitrary $I>1$.  
Note that if $\tau$ is decreased from either the tonic or phasic
regimes (see Eq.~(\ref{Tdef}) and Eq.~(\ref{Pdef})), Eq.~(\ref{Fdef})
is still satisfied, implying that the network can enter the fast
regime from either the phasic or tonic regimes.  This occurs because
the relationship between $g$ and $I$ distinguishes the latter two
regimes, while the condition for the fast regime is primarily
dependent on $\tau$.

For $\tau$ near to 1, Eq.~(\ref{T3}) is no longer valid since we
cannot ignore $e^{-T/\tau}$ with respect to $e^{-T}$.  By going back
to (\ref{trans}) for $\tau\sim 1$ and $T>>1$
we get the new period relation
\begin{equation}
1 \simeq I(1-e^{-T}) - g T e^{-T}
\end{equation}
which again can be satisfied (since $T>>\tau\sim 1$) for $I\sim 1$ and
arbitrary $g$ or for $g$ large 
enough and arbitrary $I>1$.

The period is identical for nonsaturating synapses since the synapse
is so fast it does not have a chance to reach saturation.
The period also does not
change much for model 1.  This is because the synapse
decays quickly (\ie~$\tau<<T$).  Thus over most of the period,
$\mu(t)\sim t$ is a good approximation to Eq. (\ref{mu}).
  
\section{Relationship between Full and Reduced Models:  Two
Examples}\label{example} 

The three regimes analyzed above have been observed in numerical
simulations of conductance-based models.  The phasic and tonic regimes
were reported in \citeauthor{white97}~(\citeyear{white97}) and the
fast regime, for spiking and/or bursting neurons, has been observed
previously~\cite{friesen94,perkel74,skinner94,vrees94,wang92}. Here,
we show that a stronger correspondence can be drawn between the
reduced and the conductance-based models.  We show that the period for
the reduced model as a function of the parameters
$\tilde{\tau}$, $\tilde{I}$, and $\tilde{g}$ approximates the period
obtained from simulations of the two conductance-based neurons given
in the Appendix.  Specifically we show that the period data
$\tilde{T}=\tilde{T}(\tilde{I},\tilde{g},\tilde{\tau})$ obtained from
the simulations of the conductance based models is well approximated
by the period function for the reduced models obtained from the
relation (\ref{transfull}).

The period relation (\ref{transfull}) for the reduced model is in
terms of dimensionless quantities $T$, $I$, $g$, and $\tau$
(\ie~$T=T(I,g,\tau)$) .  To compare to the numerical results we
restore the 
dimensions using the scaling transformations given in
(\ref{transform}), namely
\begin{equation}
\tilde{T}=\frac{T}{\tau_m}, \qquad I=\frac{\tilde{I}+ I_r}{I_T},
\qquad g= \frac{\tilde{g}}{g_T}.
\label{invtrans}
\end{equation}
Our task therefore is to find a single set of scaling parameters
$I_r$, $I_T$, $g_T$ and $\tau_m$ for which the reduced model period
matches that of the conductance-based model.

We first numerically generated (for each of the two models given in
the appendix) three tables of period data.  In the first table, we
listed $\tilde{T}$ as $\tilde{I}$ varied for several fixed values of
$\tilde{g}$ and $\tilde{\tau}$, while the other two tables similarly
listed $\tilde{T}$ as $\tilde{g}$ or $\tilde{\tau}$ varied
respectively. Each table was then considered a numerical function,
$\tilde{T}_{(\tilde{g},\tilde{\tau})}(\tilde{I})$,
$\tilde{T}_{(\tilde{I},\tilde{\tau})}(\tilde{g})$ or
$\tilde{T}_{(\tilde{I},\tilde{g})}(\tilde{\tau})$, and we looked
for a choice of scaling parameters (\ref{invtrans}) such that the
reduced model period function $T=T(I,g,\tau)$ implicit in
(\ref{transfull}) was closest to the
numerical functions.  We defined a measure of error as the sum of the
absolute values of the maximal deviation between the reduced and
conductance-based models (\ie~${\rm err}=\sum \max | \tilde{T} - T
|$, where the maximum was taken over all period values within a table
and the sum was taken over the three tables).  The fit error between
the conductance-based and the reduced model was then minimized using
the function FMINS in the software package MATLAB. Critical to the
success of the fitting procedure was finding a good initial guess.
This was obtained by adjusting the scaling parameters by hand and
fitting the functions by eye.

\begin{figure}
\centerline{\epsfig{figure=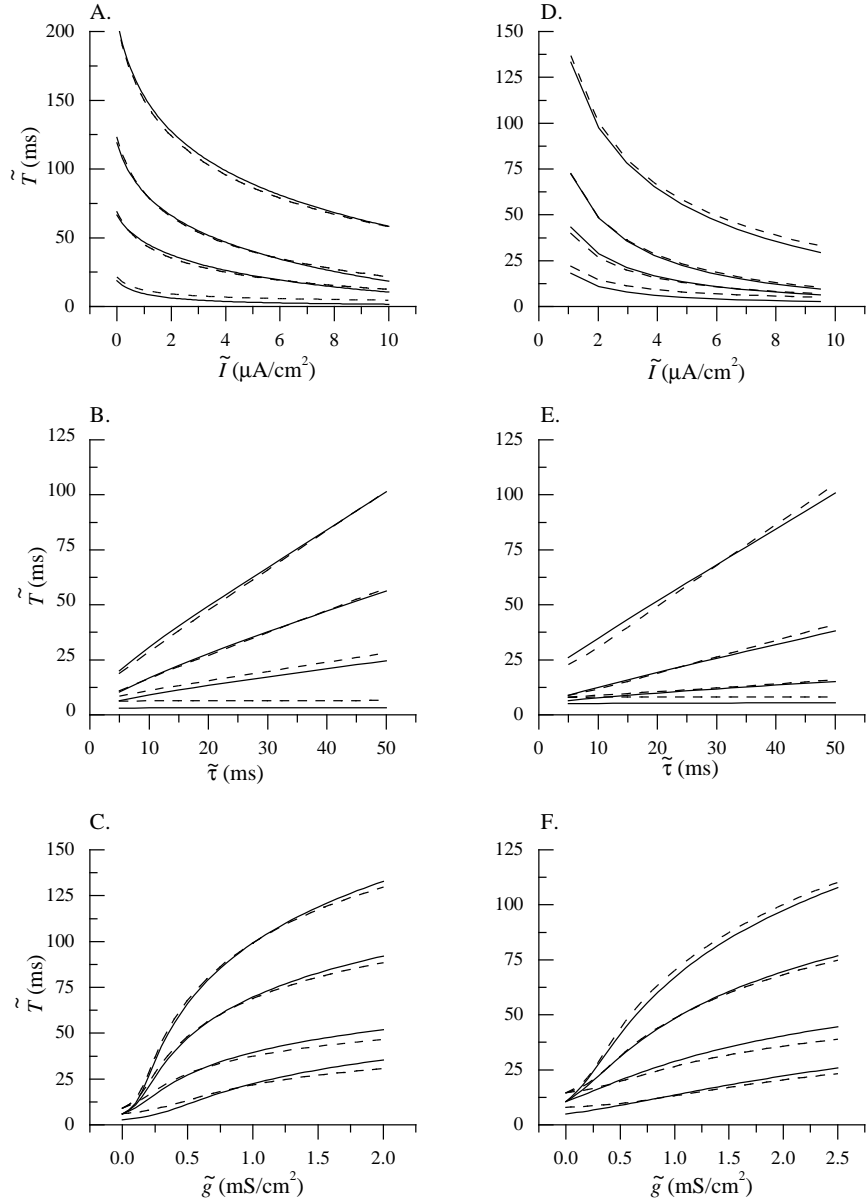, height=7.in,bbllx=14pt,bblly=9pt,bburx=597pt,bbury=784pt, clip=}}
\caption{Shown are example slices of the numerically determined period
for the conductance-based models (dashed lines) compared to the
period of the reduced model
obtained from Eq.~(\ref{transfull}) (solid lines).  Figures A), B)
and C) show results for the White \et (1998) model with memory
coefficient $a=0.30$, and scaling parameters $I_r=1.9155$, $I_T=1.4337$,
$\tau_m=12.0230$, and $g_T=0.0851$.  Figures D), E) and F) show
results for the 
reduced Traub and Miles neuron with $a=0.74$, $I_r=1.3546$, $I_T=1.6211$,
$\tau_m=16.1158$, and $g_T=0.1111$.  From top to bottom the slices
have parameters: 
A) $(\tilde{g},\tilde{\tau})=(2,50), (1,32.5), (1,15), (0.05,5)$ 
B) $(\tilde{I},\tilde{g})=(1.64,1),(5,1),(5,0.5),(5,0.05)$, 
C) $(\tilde{I},\tilde{\tau})=(1.74,50),(1.74,32.5),(1.74,15),(5,15)$, 
D) $(\tilde{g},\tilde{\tau})=(2,50),(1,32),(1,15),(0.5,5)$
E) $(\tilde{I},\tilde{g})=(1.085,1),(4.825,1.25),(4.825,0.5),(4.825,0.05)$
and F) $(\tilde{I},\tilde{\tau})=(2.02,50),(2.02,32.5),(2.02,15),(4.825,15)$.
The phasic and tonic regimes are most clearly distinguished in B) and E)
where the period is proportional to $\tilde{\tau}$ in the upper curves
and independent of $\tau$ in the lower curves.} 
\label{fig:freq1}
\end{figure}

The memory function $a=\exp(\tilde{\alpha}\Delta\tilde{t})$ was fixed
for each conductance-based model.  $\Delta \tilde t$ was defined as the spike
width at $V=0$, while both models used $\tilde{\alpha}=1$.  The
results of the fits for both models are shown in Fig.~\ref{fig:freq1}.
Note that each neuron model has a single transformation set which is
used in all three panels.  The phasic and tonic regimes are evident in
the figures.  We did not explore the fast regime.  The period derived
from the reduced model (\ref{transfull}) is seen to capture quite well
the dependence of the period upon the parameters of the
conductance-based models.  Note that the White~\et~model has a
different spike shape from the reduced Traub and Miles model (see
Fig.~\ref{fig:spike}).  The latter has a significant negative
overshoot while the former does not.  Nevertheless, the quality of the
fit is equally good in both cases.

The fit tends to work best in the phasic regime.  This is expected
because the period behavior is dominated by the choice of synaptic
model (which is the same in the analysis and simulations).  In the
tonic regime, the effect of the synapse is weak and the frequency
characteristics are more like that of the uncoupled neuron.  Thus
differences in the intrinsic membrane dynamics will be the most
prominent here.  The period of the uncoupled reduced model obeys the
classic result $T= \ln(I/I-I^*)$~\cite{tuckwell}, and we showed in the
tonic regime that the synapse affects the period only by changing the
amount of tonic input the neuron receives.  The period of the Type I
neurons in the sense of \citeauthor{bard96}~(\citeyear{bard96}) behave
as $T=(I-I^*)^{-1/2}$ near the bifurcation point.  Thus at least near
the onset of firing, the frequency behavior as a function of $I$
should not be the same.

As the synapse becomes stronger, the influence of the synaptic
dynamics begins to compete with the intrinsic membrane dynamics.  Deep
within the phasic regime the synaptic dynamics completely dominate and
the period is primarily determined by the response of the neuron to
inhibition.  Our analysis showed that the intrinsic dynamics are not
important in the phasic regime.  However, the use of model 2 instead
of model 1 for equation (\ref{transfull}) will result in quantitative
inaccuracies even well within in the phasic regime.

\section{Discussion and Conclusions}

In this paper, we show that a fully synchronized network can have its
frequency determined by several parameters, including applied
currents, synaptic strength and synaptic decay time.  However, in
regimes other than the phasic regime, in which parameters other than
the synaptic decay dominate the network frequency, even mild
heterogeneity can eliminate coherence~\cite{white97}.  Indeed, in that
work we showed that increasing the drive $I$, which increases the
uncoupled frequency of each cell or the common frequency of a
homogenous inhibitory network, can desynchronize a mildly
heterogeneous network.  This is true even if the percentage difference
in natural frequencies over the population is held constant.
Decreasing the drive sufficiently puts the network in a regime where
inhibition suppresses spiking in less excitable cells.  This also
causes a loss of coherence albeit through a different mechanism.
These two effects limit the possible frequencies at which the network
can be synchronized, and the time scale of the synapse plays a crucial
role in determining the width of this frequency band.  

\begin{figure}
\centerline{\epsfig{figure=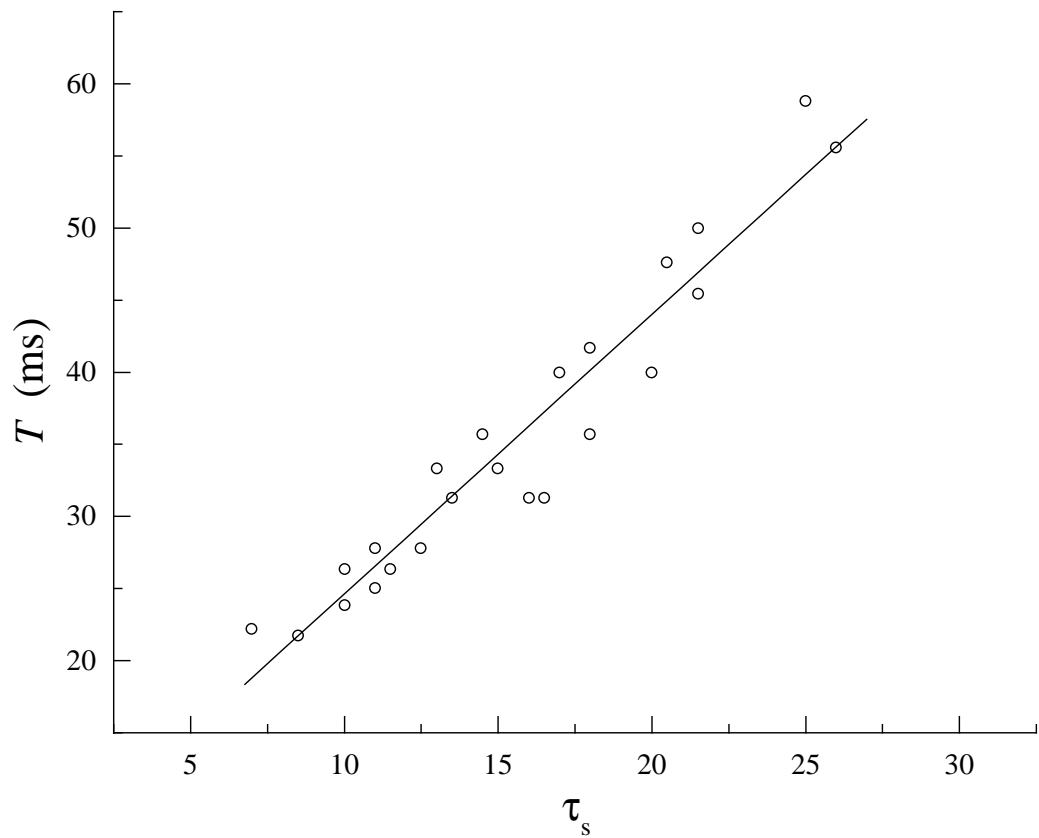,height=8.in,bbllx=14pt,bblly=9pt,bburx=597pt,bbury=784pt,clip=}}
\caption{Plot of period vs. synaptic decay time for a hippocampal
slice showing linear behavior indicative of the phasic regime.
Responses were recorded from inhibitory neurons evoked by glutamate
application during a wash-in of 2 $\mu$M
pentobarbital~(Whittington~\et, 1995).
Experimental data is provided courtesy of J. Jefferys.}
\label{fig:jeff}
\end{figure}

With respect to the hippocampus, the analysis
predicts the fast 200 Hz 
ripples observed in CA1~\cite{ylinen95}, cannot be solely mediated by
GABA$_A$ synaptic inhibition, whose time scales are much slower than
5~ms. An additional mechanism such as
gap junction mediated electrical coupling may be responsible for
the observed synchronous fast ripples.
The 40 Hz gamma rhythm, which is on the order of the time
scale of GABA$_A$, can be supported by inhibition alone.  This was
observed in experiments in hippocampal
slices~\cite{whit95,traub96,jeff96}.  Our results on
the dependence of the period on the synaptic decay time constant
clarifies the experimental
observations. Our analysis predicts that for 
coherent oscillations, the network should be in the phasic regime.
To compare with the analysis, we replotted the experimental data of
\citeauthor{whit95}~(\citeyear{whit95}) as {\em
period} versus synaptic decay time in Fig.~\ref{fig:jeff}.  We note
that the period
$T$ is proportional to the decay time $\tau$, as predicted for the
phasic regime.  Furthermore, $\tau / T$ has a value less than 1,
consistent with the requirement for being in the phasic regime.

For homogeneous networks,  the time scale of
the inhibition is not critical for the frequency of a synchronized
network.  A homogeneous network can stably synchronize at parameter
regimes outside the phasic regime.  However, for a mildly
heterogeneous network, the synchronization mechanism must not only
draw together phases, but must also help create a common frequency.
As shown in~\cite{chow97}, inhibitory synapses in the phasic regime
{\em do} provide such a mechanism for heterogeneous networks.
We wish to point out that this frequency dependent synchronization is
not necessary for all networks.  For instance,   
consider a network coupled with (nonrectified) electrical coupling.
In the fully synchronized state, 
the coupling currents disappear and hence do not contribute to the
network frequency.  With mild heterogeneity, the network still can maintain
coherence, but there is no time constant to 
affect the network frequency.

The connection of the frequency of a synchronized network with
inhibition to the inhibitory decay time has previously been understood
intuitively~\cite{destexhe93,skinner93,kopell,wang96}.  Here we
present a simple example to clarify how all of the relevant time
scales interact to produce the regimes in which that intuition is
correct.  These conclusions assume that the membrane potential between
spikes is governed by a passive decay process that approaches a fixed
threshold.  The reduction from model 1 to model 2 simplifies the
effect of the synapse.  The accuracy of our analytical
calculations depend on the validity of these approximations.
Nevertheless, we expect the three regimes to still exist for a much
larger class of conductance based models and models of synapses.
Evidence for this conjecture was given in Section~\ref{example}.
We note that~\citeauthor{IH97}~(\citeyear{IH97}) have shown that for weakly
coupled systems an arbitrary Type I excitable neuron can be
transformed into integrate-and-fire form by a piece-wise continuous change
of variables.  Our results show that this correspondence can hold
beyond the weak coupling limit.

\acknowledgements
We wish to thank B. Ermentrout for many clarifying discussions and
help in computing phase response curves.
We also thank  S. Epstein, W. Gerstner, C. Linster, 
J. Rinzel, C. Soto-Trevi\~no, and R. Traub,  for helpful discussions.
We thank J.~Jefferys for providing us with experimental data.
This work was supported in part by 
the National Science Foundation (DMS-9631755 to N.K. and J.W.), 
the National Institutes of Health (MH47150 to N.K.; 
1R29NS34425 to J.W.), 
and The Whitaker Foundation (to J.W.)

\appendix

\section{Neuron Dynamics}\label{app:neu}

For our physiologically based neuron we consider a
single-compartment model
with inhibitory synapses obeying first-order kinetics.  
The membrane potential obeys the current balance
equation
\begin{equation}
C\frac{dV_i}{dt}=\tilde{I}-I_{Na}-I_K-I_L-I_s,
\label{membrane}
\end{equation}
where $C=1 \mu$F/cm$^2$, $\tilde{I}$ is the applied current,
$I_{Na}=g_{Na} m^3 h (V_i - V_{Na})$ and $I_K=g_Kn^4(V_i-V_K)$ are
the spike generating currents, $I_L= g_L(V_i-V_L)$
is the leak current and $I_s = \sum_j^N \tilde{g} S_j(t) (V_i-V_s)$ is the
synaptic current. 

The interneuron model in White \et~(\citeyear{white97}) used parameters:
$g_{Na}=30$~mS/cm$^2$, $g_K=20$~mS/cm$^2$, 
$g_L=0.1$~mS/cm$^2$,
$V_{Na}=45$~mV, $V_K=-80$~mV, $V_L=-60$~mV, $V_s=-75$~mV.
The activation variable $m$ was assumed fast and substituted with
its asymptotic value 
$m=m_\infty(v)=(1 + \exp[-0.08(v+26)])^{-1}$.  The gating variables $h$
and $n$ obey
\begin{equation}
\frac{dh}{dt}=\frac{h_\infty(v)-h}{\tau_h(v)},
\qquad
\frac{dn}{dt}=\frac{n_\infty(v)-n}{\tau_n(v)},
\end{equation}
with 
$h_\infty(v)=(1 + \exp[0.13(v+38)])^{-1}$, 
$\tau_h(v)=0.6/(1 + \exp[-0.12(v+67)])$,
$n_\infty(v)=(1 + \exp[-0.045(v+10)])^{-1}$, and
$\tau_n(v)=0.5+2.0/(1+\exp[0.045(v-50)])$.

The reduced Traub and Miles
model~\cite{ermenkopell98,traub96b} used 
parameters: $g_{Na}=100$~mS/cm$^2$, 
$g_K=80$~mS/cm$^2$,  $g_L=0.1$~mS/cm$^2$,
$V_{Na}=50$~mV, $V_K=-100$~mV, $V_L=-67$~mV, $V_s=-80$~mV;
$m=m_\infty(v)=\tilde{\alpha}_m(v)/(\tilde{\alpha}_m(v)+\tilde{\beta}_m(v))$,
where 
$\tilde{\alpha}_m(v)=0.32 (54+v)/(1-\exp(-(v+54)/4))$ and 
$\tilde{\beta}_m(v)=0.28 (v+27)/(\exp((v+27)/5)-1)$;
\begin{equation}
\frac{dn}{dt}=\tilde{\alpha}_n(v)(1-n)-\tilde{\beta}_n(v)n
\end{equation}
with $\tilde{\alpha}_n(v)=0.032(v+52)/(1-\exp(-(v+52)/5))$, 
$\tilde{\beta}_n(v)=0.5\exp(-(v+57)/40)$;
$h=h_\infty(v)=\max[1 - 1.25 n, 0]$.

For both models,
the synaptic gating variable $S_j(t)$ is assumed to obey first order
kinetics of the form
\begin{equation}
\frac{dS_j}{dt} = \tilde{{\alpha}} F(V_j)(1-S_j)-\tilde{{\beta}} S_j,
\end{equation}
where $\tilde{{\alpha}}=1$ and $\tilde{{\beta}}$ are
respectively the 
synaptic rise and decay rates and $F(V_i)=1/(1+\exp[-V_i])$.
Transmission delays are neglected.  We are interested in the response
of the system to changes in the applied current $\tilde{I}$, synaptic
conductance $\tilde{g}$, and synaptic decay time $\tilde{\tau}\equiv
\tilde{{\beta}}^{-1}$.  The ODEs were integrated using either a fourth-order
Runge-Kutta method or the Gear method.

\end{document}